\documentclass[12pt]{article}
\usepackage{verbatim,graphicx,amssymb,cite}
\newcommand{\be}{\begin{equation}}
\newcommand{\ee}{\end{equation}}

\renewcommand{\thefootnote}{\fnsymbol{footnote}}
\begin{document}
\setcounter{page}{0}
\begin{titlepage}
\vspace*{-2.0cm}
\begin{flushright}
arXiv: 1007.2167 [hep-ph]
\end{flushright}
\vspace*{0.1cm}
\begin{center}
{\Large \bf Improving LMA predictions with non-standard interactions: 
neutrino decay in solar matter?}\\
\vspace{1.0cm}

{\large
C. R. Das\footnote{E-mail: crdas@cftp.ist.utl.pt},
Jo\~{a}o Pulido\footnote{E-mail: pulido@cftp.ist.utl.pt}\\
\vspace{0.15cm}
{{\small \sl CENTRO DE F\'{I}SICA TE\'{O}RICA DE PART\'{I}CULAS (CFTP)\\
 Departamento de F\'\i sica, Instituto Superior T\'ecnico\\
Av. Rovisco Pais, P-1049-001 Lisboa, Portugal}\\
}}
\vspace{0.25cm}
\end{center}
\vglue 0.6truecm

\begin{abstract}
It has been known for some time that the well established LMA solution to the observed 
solar neutrino deficit fails to predict a flat energy spectrum for SuperKamiokande as 
opposed to what the data indicates. It also leads to a Chlorine rate which appears to 
be too high as compared to the data. We investigate the possible solution to these
inconsistencies with non standard neutrino interactions, assuming that they come as 
extra contributions to the $\nu_{\alpha}\nu_{\beta}$ and $\nu_{\alpha}e$ vertices that 
affect both the propagation of neutrinos in the sun and their detection. We find
that, among the many possibilities for non standard couplings, only the diagonal
imaginary ones lead to a solution to the tension between the LMA predictions and the data,
implying neutrino instability in the solar matter. Unitarity requirements further restrict 
the solution and a neutrino decay into an antineutrino and a majoron within the sun 
is the one favoured. Antineutrino probability is however too small to open the possibility
of experimentally observing antineutrinos from the sun due to NSI.
\end{abstract}

\end{titlepage}

\setcounter{footnote}{0}
\renewcommand{\thefootnote}{\arabic{footnote}}
\section{Introduction}
\label{sec1}

Neutrino non-standard interactions (NSI) have been introduced long ago \cite{Guzzo:1991hi,
Roulet:1991sm} to account for a possible alternative solution to the solar neutrino 
problem. Since then a great deal of effort has been dedicated to study its possible
consequences. To this end possible NSI signatures in neutrino processes have been investigated, 
models for neutrino NSI have been developed and bounds have been derived \cite{Grossman:1995wx,
Johnson:1999ci,Datta:2000ci,Huber:2001zw,Huber:2001de,Ota:2001pw,Huber:2002bi,Davidson:2003ha,
Barranco:2005ps,Mangano:2006ar,Antusch:2008tz,Biggio:2009kv,Gago:2009ij,Biggio:2009nt,Wei:2010ww}.
Specific investigations of NSI in matter have also been performed within the context
of supernova \cite{EstebanPretel:2007yu} and solar neutrinos \cite{Berezhiani:2001rt,
Friedland:2004pp,Guzzo:2004ue,Miranda:2004nb,Bolanos:2008km,Escrihuela:2009up}.

Although LMA is generally accepted as the dominant solution to the solar neutrino problem 
\cite{Fogli:2008ig,Schwetz:2008er}, not only its robustness has been challenged by NSI, as it 
can shift the LMA solution to the dark side region of parameter space \cite{Miranda:2004nb},
but also some inconsistencies remain regarding its agreement with the data \cite{Das:2009kw,
Pulido:2009sb}. In fact, while the SuperKamiokande (SK) energy spectrum appears to be flat
\cite{Fukuda:2002pe,:2008zn}, the LMA prediction shows a clear negative slope in the same energy 
range. With the expected improvement in the trigger efficiency for threshold electron energies 
as low as $3~MeV$ to be reached in the near future \cite{Smy}, such a disagreement, if it 
persists, may become critical. Moreover the LMA solution predicts an event rate for the 
Cl experiment \cite{Cleveland:1998nv} which is 2$\sigma$ above the observed one 
\cite{deHolanda:2003tx}. These are motivations to consider `beyond LMA' solar neutrino 
solutions in which NSI may play a subdominant, although important role.

In order for NSI to be detectable and therefore relevant in physical processes, the 
characteristic scale of the new physics must not be too much higher than the scale of
the physics giving rise to the Standard Model interactions, $\Lambda_{EW}\simeq G_F^{-1/2}$. 
Possible realisations are one loop radiative models of Majorana neutrino mass 
\cite{AristizabalSierra:2007nf}, supersymmetric SO(10) with broken D-parity 
\cite{Malinsky:2005bi}, the inverse seesaw in a supersymmetry context 
\cite{Bazzocchi:2009kc} or triplet seesaw models \cite{Malinsky:2008qn}. 
Since the scale at which the new interaction arises is supposed to be not too far 
from the electroweak scale, its coupling may be parameterised by $G_F\varepsilon$ 
where $\varepsilon\simeq \Lambda^2_{EW}/\Lambda^2_{NP}\simeq 10^{-2}$ for d=6 or 
$\varepsilon\simeq \Lambda^4_{EW}/\Lambda^4_{NP}\simeq 10^{-4}$ for d=8 operators 
respectively. For type I seesaw, NSI are of course negligible.

In this paper we will be concerned with NSI both at the level of propagation
through solar matter and at the level of detection. Matter NSI are defined through 
the addition of an effective operator to the Lagrangian density   
\begin{equation}
{\cal L}^{M}_{NSI}=-2\sqrt{2}G_F \varepsilon^{fP}_{\alpha \beta}[\bar f\gamma^{\mu}
P f][\bar\nu_{\alpha}\gamma_{\mu}P_L \nu_{\beta}]
\label{eq1}
\end{equation}
where $f=e,u,d$, $P$ denotes the projection operator for left and right chirality 
and $\varepsilon^{fP}_{\alpha \beta}$ parameterizes the deviation from 
the standard interactions. At present there is no evidence at all of such operators
generated at a scale $\Lambda_{NP}$, hence the variety of theoretical models for the 
physics accessible to the LHC. 

There are several ways to introduce NSI. For instance in fermionic seesaw models, once 
the heavy fermions (singlets or triplets) are integrated out, modified couplings of 
leptons to gauge bosons are obtained in the form of a non-unitary leptonic matrix.
The strong bounds on the deviation from the unitarity of this matrix constrain these
NSI to be $\lesssim O(10^{-3})$ \cite{Antusch:2008tz}. Alternatively NSI can be generated 
by other new physics above the electroweak scale not related to neutrino masses. As a 
consequence an SU(2) gauge invariant formulation of NSI is required, since any gauge 
theory beyond the standard model must necessarily respect its gauge symmetry. 
Strong bounds from four charged fermionic processes \cite{Berezhiani:2001rs} and 
electroweak precision tests requiring fine tunings imply that possibilities are 
limited for such scenarios \cite{Biggio:2009kv,Gavela:2008ra}. Another way to introduce
NSI is by assuming that there are extra contributions to the vertices $\nu_{\alpha}
\nu_{\beta}$ and $\nu_{\alpha}e$. In such a case the parameters $\varepsilon_{\alpha
\beta}$ describe the deviation from the standard model vertices and are treated like
the standard interactions. It is possible that other effects are present in this case
that depend on the nature and number of particles that may be introduced in a particular
model. We adopt this procedure in the present paper and assume these model dependent 
effects to be negligible.

The paper is organized as follows: section \ref{sec2} is devoted to the study of the propagation 
and detection of solar neutrinos. We start by reviewing the derivation of the neutrino 
refraction indexes with standard interactions (SI) and their generalization to NSI in 
order to obtain the matter 
Hamiltonian. The survival and conversion probabilities to $\nu_{\mu}$ and $\nu_{\tau}$
are then evaluated through the numerical integration of the Schr\"{o}dinger like equation
using the Runge-Kutta method and the experimental event rates are obtained. We use the
reference solar model with high metalicity, BPS08(GS) \cite{PenaGaray:2008qe}. In section \ref{sec3} 
we investigate the influence of the NSI couplings on these rates in order to find whether
and how the fits can be improved with respect to the LMA ones. We concentrate in 
particular on the elimination of the upturn in the SuperKamiokande spectrum predicted by 
LMA for energies below 8-10 $MeV$ not supported by the data \cite{Das:2009kw,Pulido:2009sb,
Fukuda:2002pe,:2008zn} and on the Chlorine rate whose LMA prediction exceeds the data by 
2$\sigma$ \cite{Cleveland:1998nv,deHolanda:2003tx}. We find the intriguing result that
only imaginary diagonal couplings in the matter Hamiltonian provide a suitable solution 
to the problem. In fact all other possibilities, namely imaginary off diagonal couplings 
or real couplings, either diagonal or off diagonal, are unable to change the LMA solution.
This means that eliminating the tension with the data implies complex matter Hamiltonian 
eigenvalues and therefore neutrino decay induced by the solar matter. In section  \ref{sec4}
it will be seen that the condition of consistency with the data on the event rates is 
not sufficient, owing to the requirement of unitarity which severely restricts the correct 
solution. Furthermore, the neutrino decay in the sun suggested long ago 
into an antineutrino and a scalar particle (majoron) \cite{Nussinov:1987pc,
Raghavan:1987uh,Berezhiani:1987gf,Berezhiani:1993iy} is the favoured 
channel. In section \ref{sec5} we present a brief discussion and summarize our main conclusions.

\section{Interaction potentials, the Hamiltonian and the rates}
\label{sec2}

In this section we develop the framework that will be used as the starting point for the
analysis of the NSI couplings in section  \ref{sec3}. To this end we review the derivation of the 
neutrino interaction potentials in solar matter, its generalization to non standard 
interactions along with the corresponding matter Hamiltonian and the event rates.

\subsection{Interaction potentials and the Hamiltonian}
\label{subsec2.1}

While $\nu_e$'s propagate through solar matter their interaction with electrons proceeds
both through charged and neutral currents (CC) and (NC). Recalling that for standard 
interactions (SI) each tree level vertex accounts for a factor
\begin{equation}
\frac{g_L}{cos\theta_W}(T_{3L}-2Q_f sin^2\theta_{W}),
\label{eq2}
\end{equation}
inserting the W, Z propagators and the electron external lines, one gets for the $\nu_e$ 
interaction potentials
\begin{equation}
(V_e)_{CC}=G_F\sqrt{2}N_e~,~(V_e)_{NC}=G_F/\sqrt{2}(-1+4sin^2\theta_W)N_e
\label{eq3}
\end{equation}
where $G_F$ is the Fermi constant, $G_F/\sqrt{2}=g_L^2/8m^2_W$. For the interactions
with quarks only neutral currents are involved and the additivity of the quark-current 
vertices gives for protons
\begin{equation}
V_p=(V_p)_{NC}=G_F/\sqrt{2}(1-4 sin^2 \theta_W)N_e
\label{eq4}
\end{equation}
and for neutrons
\be
V_n=(V_n)_{NC}=-G_F/\sqrt{2}N_n.
\label{eq5}
\ee
Hence the neutrino interaction potential is for standard interactions\footnote{All expressions
are divided by 2 to account for the fact that the medium is unpolarized.}
\be
V(SI)=V_e+V_p+V_n=G_F\sqrt{2}N_e\left(1-\frac{N_n}{2N_e}\right)=
V_c+V_n
\label{eq6}
\ee
with $V_e=(V_e)_{CC}+(V_e)_{NC}$ and $V_c=V_e+V_p=G_F\sqrt{2}N_e$.

In order to introduce NSI we assume that each diagram associated to neutrino
propagation in matter (i.e. CC and NC currents in $\nu_{\alpha}~e^{-}$ and NC
currents in $\nu_{\alpha}~u,~\nu_{\alpha}~d$ scattering) is multiplied by a factor
$\varepsilon_{\alpha\beta}^{e,u,d~P}$ parameterising the deviation from the standard model.
So we assume that the interaction potential for $\nu_{\alpha}$ ($\alpha=e,~\mu,~\tau)$ on 
electrons involves both CC and NC giving rise to possible lepton flavour violation: 
$\nu_{\alpha}$ for $\alpha \neq e$ {\it may} have CC. So for the charged current of $\nu_e$
with electrons we have
\be
(V_{e})_{CC}(NSI)=\frac{g^2_L}{2m^2_W}(\varepsilon_{\alpha \beta}^{eP})_{CC}N_e
\label{eq7}
\ee
and for the neutral current
\be
(V_{e})_{NC}(NSI)=\frac{g^2_L}{4m^2_W}(-1+4sin^2\theta_W)(\varepsilon_{\alpha \beta}^{eP})_{NC}N_e
\label{eq8}
\ee
where the NSI couplings affecting the CC and NC processes should in principle be distinguished.

Using equation (\ref{eq2}) and the additivity of the quark-current vertices, one gets for the 
neutrino interaction potential with protons 
\be
V_p(NSI)=\frac{g^2_L}{2m^2_W}\left[\varepsilon_{\alpha \beta}^{uP}-
\frac{\varepsilon_{\alpha \beta}^{dP}}{2}-\left(\frac{4}{3}2\varepsilon_{\alpha \beta}^{uP}-
\frac{2}{3}2\varepsilon_{\alpha \beta}^{dP}\right)sin^2\theta_W\right]N_e.
\label{eq9}
\ee
Similarly for neutrons
\be
V_n(NSI)=\frac{g^2_L}{2m^2_W}\left(\frac{\varepsilon_{\alpha \beta}^{uP}}{2}-
\varepsilon_{\alpha \beta}^{dP}\right)N_n.
\label{eq10}
\ee
In both (\ref{eq9}) and (\ref{eq10}) only neutral currents are involved.

Adding (\ref{eq7}), (\ref{eq8}), (\ref{eq9}) and (\ref{eq10}) and dividing by 2 one finally gets
\begin{eqnarray}
V(NSI) & = & G_F\sqrt{2}N_e\left[(\varepsilon_{\alpha \beta}^{eP})_{CC}+
\left(-\frac{1}{2}+2sin^2\theta_W \right)
(\varepsilon_{\alpha \beta}^{eP})_{NC}+\left(1-\frac{8}{3}sin^2\theta_W+\frac{N_n}{2N_e} \right) 
\varepsilon_{\alpha \beta}^{uP}\right.\nonumber\\
&& + \left.\left(-\frac{1}{2}+\frac{2}{3}sin^2\theta_W-\frac{N_n}{N_e} \right)
\varepsilon_{\alpha \beta}^{dP} \right]
\label{eq11}
\end{eqnarray}
so that the full interaction potential is the sum of eqs.(\ref{eq6}) and (\ref{eq11}). 

In the case of the standard interactions, the interaction potentials for $\nu_e$ and 
$\nu_{\alpha}$ constitute a diagonal matrix because they cannot be responsible for flavour 
change [eq.(\ref{eq6})]. This may occur as a consequence of the vacuum mixing angle
(oscillations) \cite{Fogli:2008ig,Schwetz:2008er} or the magnetic moment for 
instance \cite{Das:2009kw,Pulido:2009sb}. On the other hand, in the case of NSI 
the interaction potentials [eq.(\ref{eq11})] constitute a full matrix in neutrino flavour space. 

In order to obtain the matter Hamiltonian eqs.(\ref{eq6}) and (\ref{eq11}) must now be added. In the 
flavour basis this is
\be
{\cal H}_M=V_c \left(\begin{array}{ccc} 1 & 0 & 0\\
0 & 0 & 0\\
0 & 0 & 0\\ \end{array}\right)+
\left(\begin{array}{ccc} v_{ee}(NSI) & v_{e\mu}(NSI) & v_{e\tau}(NSI)\\
v_{\mu e}(NSI) & v_{\mu\mu}(NSI) & v_{\mu\tau}(NSI)\\
v_{\tau e}(NSI) & v_{\tau\mu}(NSI) & v_{\tau\tau}(NSI)\\ \end{array}\right)~
\label{eq12}
\ee
where in the first term, describing the standard interactions, the additive quantity $V_n$
which is proportional to the identity, has been removed from the diagonal. In the second term
$v_{\alpha\beta}$ ($\alpha,\beta=e,\mu,\tau$) denote the matrix elements of the interaction 
potential matrix (\ref{eq11}). Finally in the mass basis
\be
{\cal H}=\left(\begin{array}{ccc} 0 & 0 & 0\\
0 & \frac{\Delta m^2_{21}}{2E} &  0\\
0 &       0         &  \frac{\Delta m^2_{31}}{2E}\\ \end{array} \right)+ U^{\dagger}
{\cal H}_M U
\label{eq13}
\ee
where $U$ is the PMNS matrix \cite{Maki:1962mu}\footnote{We use the standard 
parameterization \cite{Amsler:2008zzb} for the $U$ matrix and the central value 
$sin\theta_{13}=0.13$ claimed in ref. \cite{Fogli:2008ig}.}, $E$ is the neutrino energy
and $\Delta m^2_{ij}=m_i^2-m_j^2$ with $m_i$ ($i=1,2,3$) the neutrino mass. Upon
insertion of this Hamiltonian expression in the neutrino evolution equation, the
survival ($P_{ee}$) and conversion probabilities ($P_{e\mu},P_{e\tau}$) are evaluated 
using the Runge-Kutta numerical integration.

\subsection{Neutrino electron scattering detection rates}
\label{subsec2.2}

For the detection in SuperKamiokande and SNO through $\nu_{\alpha}~e^{-} \rightarrow 
\nu_{\beta}~e^{-}$ scattering, the NSI information comes in the probabilities and the cross 
section 
\be
\frac{d\sigma}{dT}=\frac{2G_F^2 m_e}{\pi}\left[\tilde g_L^2+\tilde g_R^2\left(1-
\frac{T}{E_{\nu}}\right)^2-\tilde g_L \tilde g_R\frac{m_e T}{E^2_{\nu}}\right]
\label{eq14}
\ee
where $\tilde g_{L,R}$ are the $g_{L,R}$ couplings modified according to \cite{Barranco:2005ps}\footnote{This 
modification is different from the one in ref.\cite{Barranco:2005ps}
since each $\varepsilon$ coupling as defined in section \ref{sec2} pertains not to a single vertex 
but to a whole NSI diagram.}
\begin{eqnarray*} 
(\tilde g_{L,R})_{\nu_e}^2&=&\left|(g_{L,R})_{\nu_e}^2+\varepsilon_{ee}^{e~L,R}\right|+
\sum_{\alpha\neq e}\left|\varepsilon_{\alpha e}^{e~L,R}\right|
~~~~~~{\rm for}~~~~~~\nu_e~e^{-}
\rightarrow \nu_{\alpha}~e^{-}\\
(\tilde g_{L,R})_{\nu_\mu}^2&=&\left|(g_{L,R})_{\nu_\mu}^2+\varepsilon_{\mu\mu}^{e~L,R}\right|+
\sum_{\alpha\neq \mu}\left|\varepsilon_{\alpha\mu}^{e~L,R}\right|
~~~~~~{\rm for}~~~~~~\nu_{\mu}~e^{-}
\rightarrow \nu_{\alpha}~e^{-}\\
(\tilde g_{L,R})_{\nu_\tau}^2&=&\left|(g_{L,R})_{\nu_\tau}^2+\varepsilon_{\tau\tau}^{e~L,R}\right|+
\sum_{\alpha\neq \tau}\left|\varepsilon_{\alpha\tau}^{e~L,R}\right|
~~~~~~{\rm for}~~~~~~\nu_{\tau}~e^{-}
\rightarrow \nu_{\alpha}~e^{-}~.
\end{eqnarray*}
with $\alpha=e,\mu,\tau$.

For $\nu_e$ both charged and neutral currents are possible, so that
\be
(g_L)_{\nu_e}=\frac{1}{2}+sin^2\theta_W~,~~(g_R)_{\nu_e}=sin^2\theta_W
\label{eq15}
\ee
whereas $\nu_{\mu,\tau}$ only interact through neutral currents, hence
\be
(g_L)_{\nu_{\mu},\nu_{\tau}}=-\frac{1}{2}+sin^2\theta_W~,~~(g_R)_{\nu_{\mu},\nu_{\tau}}=
sin^2\theta_W.
\label{eq16}
\ee

These expressions are then inserted in the spectral event rate 
\begin{equation}
R^{th}_{SK,SNO}(E_e)\!\!=\!\!
\frac{\displaystyle\int_{m_e}^{{E'_e}_{max}}\!\!dE'\!\!_ef(E'_e,E_e)
\!\!\int_{E_m}^{E_M}\!\!dE\phi(E)\!\!\left[P_{ee}(E)\frac{d\sigma_{e}}{dT'}\!+\!
P_{e\mu}(E)\frac{d\sigma_{\mu}}{dT'}\!+\!P_{e\tau}(E)\frac{d\sigma_{\tau}}{dT'}\right]}
{\displaystyle\int_{m_e}^{{E'_e}_{max}}\!\!dE'\!\!_e
f(E'_e,E_e)\!\!\int_{E_m}^{E_M}\!\!dE\phi(E)\frac{d\sigma_{e}}{dT'}}
\label{eq17}
\end{equation}
which will be evaluated in the next section. Here $\phi(E)$ denotes the neutrino flux 
from Boron and hep neutrinos, $f(E^{'}_e,E_e)$ is the energy resolution function for
SuperKamiokande and SNO \cite{Fukuda:1998fd,Aharmim:2005gt} and the rest of the notation 
is standard. 

Notice that whereas the Hamiltonian (\ref{eq13}) is symmetric under the interchange 
$$\varepsilon_{\alpha\beta}^L \leftrightarrow \varepsilon_{\alpha\beta}^R~~~{\rm for}~~~e,u,d$$
such is not the case for the detection process [see eqs.(\ref{eq14})-(\ref{eq16})]. We finally note that at
the detection level the NSI couplings $\varepsilon_{\alpha\beta}^{L,R}$ are considered 
separately, as clearly seen from eq.(\ref{eq14}), whereas at the level of propagation, since the 
diagrams involved in the interaction potentials add up, their sum should instead be 
considered.

\section{NSI couplings, probabilities and spectra}
\label{sec3}

We now perform an investigation of the effect of the NSI couplings $\varepsilon_
{\alpha\beta}^{e,u,d}=|\varepsilon_{\alpha\beta}^{e,u,d}|e^{i\phi_{\alpha\beta}^{e,u,d}}$ 
on the neutrino probability and event rates. Our aim in this section is to find those 
couplings which lead to a flat SuperKamiokande spectral rate, thus improving the fit with 
respect to its LMA prediction while keeping the quality of the other solar event rate fits. 
We start by inserting one coupling at a time, first with equal CC and NC couplings, namely 
$(\varepsilon_{\alpha \beta}^{eP})_{CC}= (\varepsilon_{\alpha \beta}^{eP})_{NC}=
\varepsilon_{\alpha \beta}^{eP}$ (subsection \ref{subsec3.1}) and next with $(\varepsilon_
{\alpha\beta}^{eP})_{CC}\neq (\varepsilon_{\alpha \beta}^{eP})_{NC}$ (subsection 
\ref{subsec3.2}). We then extend our analysis to include all diagonal couplings in subsection 
\ref{subsec3.3}. It is found that complex diagonal entries in the NSI Hamiltonian are the only 
ones the modify the LMA solution.

\subsection {$(\varepsilon_{\alpha \beta}^{eP})_{CC}=(\varepsilon_{\alpha \beta}^{eP})_{NC}
=\varepsilon_{\alpha \beta}^{eP}$}
\label{subsec3.1}

For the sake of clarity we will organize the NSI couplings in three matrices according to 
whether the charged fermion in the external line is $e,~u,~d$
\be
\left(\begin{array}{ccc} \varepsilon_{ee}^{e,u,d~P} & \varepsilon_{e\mu}^{e,u,d~P} & 
\varepsilon_{e\tau}^{e,u,d~P}\\
\varepsilon^{*e,u,d~P}_{e\mu} & \varepsilon_{\mu\mu}^{e,u,d~P} & 
\varepsilon_{\mu\tau}^{e,u,d~P}\\
\varepsilon^{*e,u,d~P}_{e\tau} & \varepsilon^{*e,u,d~P}_{\mu\tau} & 
\varepsilon_{\tau\tau}^{e,u,d~P}\\ \end{array}\right).
\label{eq18}
\ee
Each set of three couplings $\varepsilon_{\alpha\beta}^{e,u,d~P}$ enters in equation
(\ref{eq11}) in the entry $v_{\alpha\beta}$ of the interaction potential matrix. Altogether 
there are 18 couplings with 36 parameters: each matrix 
of the three in eq.(\ref{eq18}) contains 6 independent entries, each with a modulus and a phase. 
We analyse one coupling at a time, by taking all others zero. We first consider the 
cases of purely real and imaginary couplings, hence 4 possibilities for each phase 
\be 
\phi_{\alpha\beta}^{e,u,d}=0,\pi/2,\pi,(3/2)\pi.
\label{eq19}
\ee
Motivated by the arguments expound in the introduction we investigate the parameter range
$|\varepsilon_{\alpha\beta}|~\epsilon~[5\times 10^{-5}~,~5\times 10^{-2}]$. We find that 

\begin{itemize}
\item Off diagonal entries $\varepsilon_{\alpha\beta}^{e,u,d~P}$ ($\alpha\neq\beta$) which
contain 3$\times$3$\times$4$=$36 possibilities for moduli and phases do not induce any
change in the LMA probability, nor any visible change in the rates, either if one or more
at a time are inserted.

\item Diagonal entries $\varepsilon_{\alpha\alpha}^{e,u,d~P}$. 
\begin{itemize}
\item[(a)] Real couplings $\varepsilon_{\alpha\alpha}^{e,u,d~P}=\pm|\varepsilon_{\alpha\alpha}^{e,u,d~P}|$
(3$\times$3$\times$2$=$18 possibilities) do not change the LMA probability, hence the rates.

\item[(b)] Imaginary couplings $\varepsilon_{\alpha\alpha}^{e,u,d~P}=
\pm i |\varepsilon_{\alpha\alpha}^{e,u,d~P}|$ (3$\times$3$\times$2$=$18 possibilities)  
lead to probabilities which diverge from $P_{LMA}$ for all $|\varepsilon_{\alpha\alpha}|>
5\times 10^{-5}$. According to the probability shape that is obtained, we group these cases in 
the following way
\end{itemize}
\end{itemize}

\begin{table}[ht]
\centering
\begin{tabular}{|c|ccc|}\hline  
  &         1                &           2               &           3\\ \hline \hline
A & $+i|\varepsilon_{ee}^{e~P}|$ & $+i|\varepsilon_{\mu\mu}^{e~P}|$ &  $-i|\varepsilon_{ee}^{e~P}|$\\
B & $+i|\varepsilon_{ee}^{u~P}|$ & $+i|\varepsilon_{\mu\mu}^{u~P}|$ &  $-i|\varepsilon_{ee}^{u~P}|$\\
C & $-i|\varepsilon_{ee}^{d~P}|$ & $-i|\varepsilon_{\mu\mu}^{d~P}|$ &  $+i|\varepsilon_{ee}^{d~P}|$\\ \hline
D & $-i|\varepsilon_{\mu\mu}^{e~P}|$ & $-i|\varepsilon_{\tau\tau}^{e~P}|$ &  $+i|\varepsilon_{\tau\tau}^{e~P}|$\\ 
E & $-i|\varepsilon_{\mu\mu}^{u~P}|$ & $-i|\varepsilon_{\tau\tau}^{u~P}|$ &  $+i|\varepsilon_{\tau\tau}^{u~P}|$\\ 
F & $+i|\varepsilon_{\mu\mu}^{d~P}|$ & $+i|\varepsilon_{\tau\tau}^{d~P}|$ &  $-i|\varepsilon_{\tau\tau}^{d~P}|$\\ \hline
\end{tabular}
\caption{\it{The NSI couplings that modify the LMA probability.}}
\label{table1}
\end{table}

\begin{figure} [htb]
\centering
\includegraphics[height=120mm,keepaspectratio=true,angle=0]{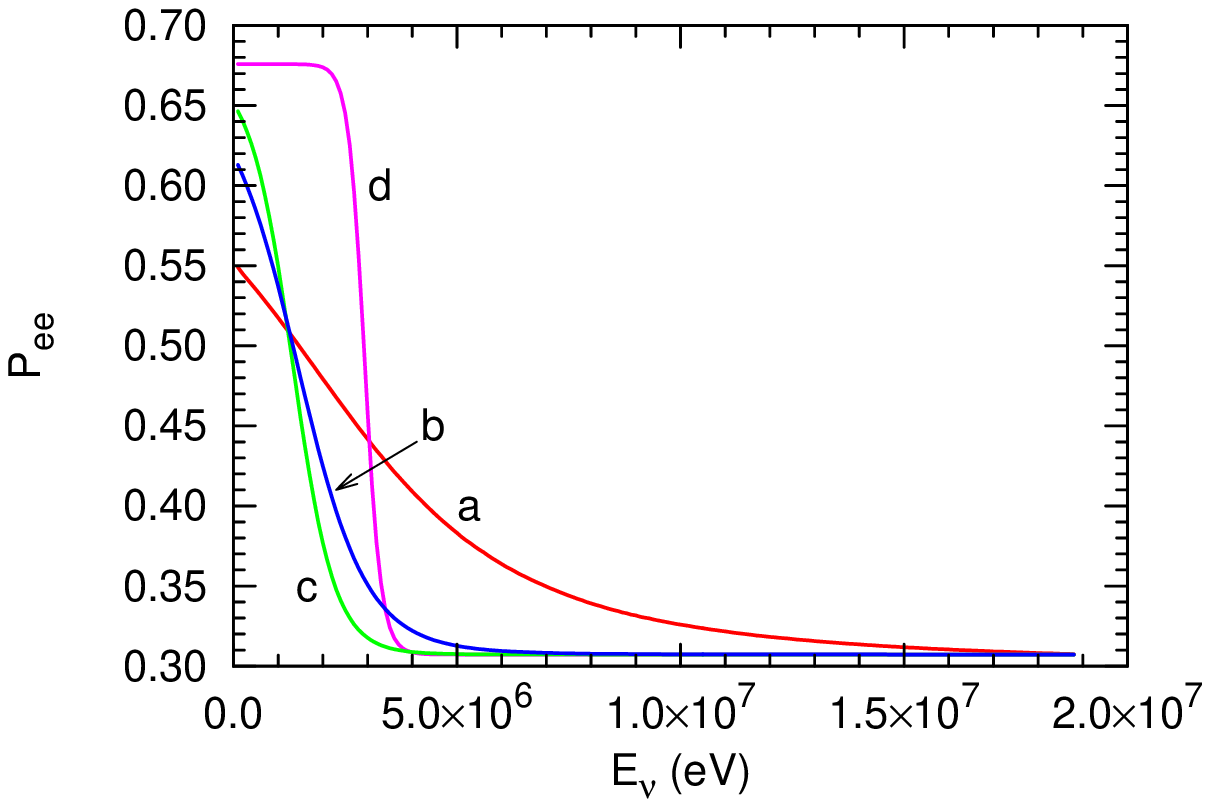}
\caption{ \it Survival probabilities as a function of neutrino energy in eV. The dashed line
is the vacuum one. The remainder are (a) the LMA one, the one providing
the best fit to the data (b) with $\varepsilon_{\mu\mu}^{e~P}=-i~1.5\times 10^{-3}$ [eq.(\ref{eq20})],
curve (c) for $\varepsilon_{\mu\mu}^{e~P}=-i~3\times 10^{-3}$ and curve (d) for
$\varepsilon_{\mu\mu}^{e~P}=+i~5\times 10^{-3}$ with other non standard couplings vanishing
in each case. The last two curves ((c) and (d)) lead to an unacceptably high Ga rate prediction
\cite{Cattadori:2005gq,Gavrin:2007wc}.}
\label{fig1}
\end{figure}

\begin{figure} [htb]
\centering
\includegraphics[height=120mm,keepaspectratio=true,angle=0]{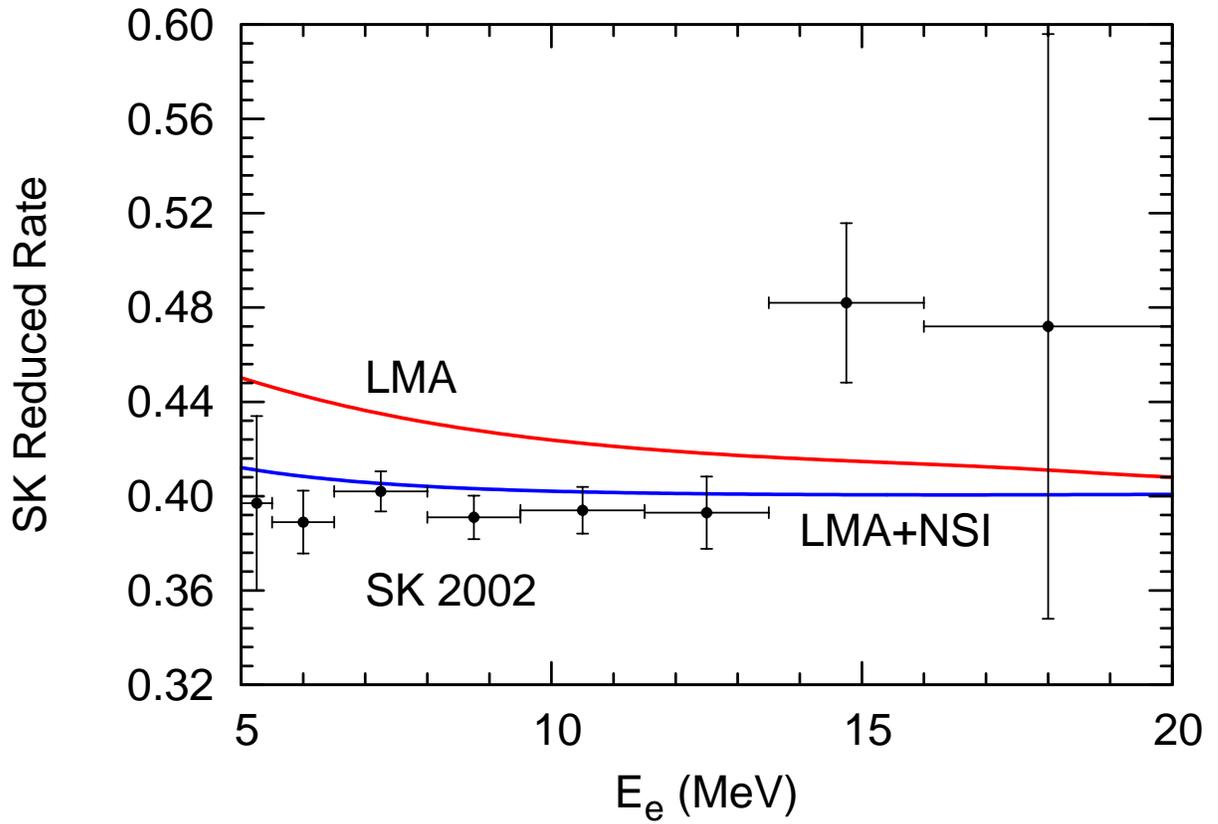}
\caption{ \it Predictions for SuperKamiokande (units in MeV for electron energy). 
The upper curve is the LMA spectrum and the lower curve is the LMA spectrum with non 
standard interactions as in eqs.(\ref{eq23}) or (\ref{eq24}). These are superimposed on the data 
published by the Collaboration in 2002 \cite{Fukuda:2002pe}.}
\label{fig2}
\end{figure}

\begin{figure} [htb]
\centering
\includegraphics[height=120mm,keepaspectratio=true,angle=0]{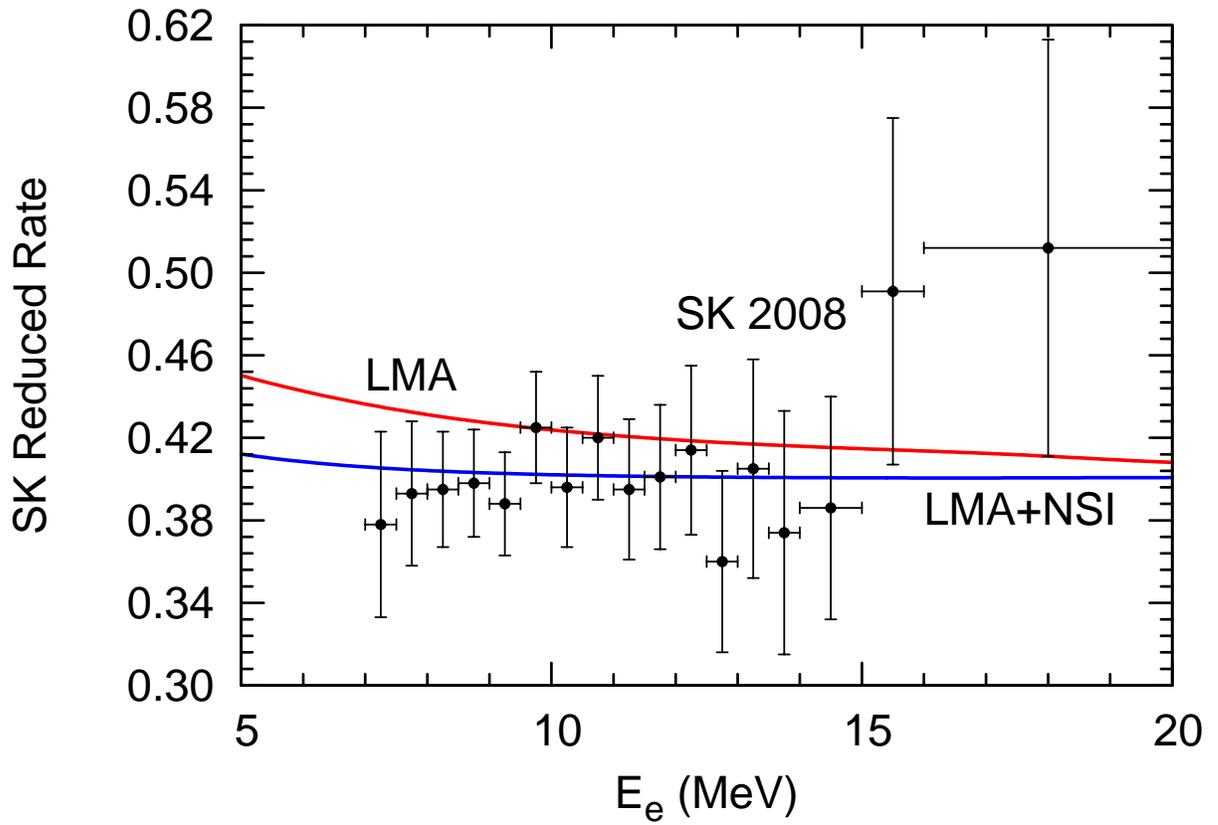}
\caption{\it Same as fig.\ref{fig2} with the data published in 2008 \cite{:2008zn}.}
\label{fig3}
\end{figure}

\begin{figure} [htb]
\centering
\includegraphics[height=120mm,keepaspectratio=true,angle=0]{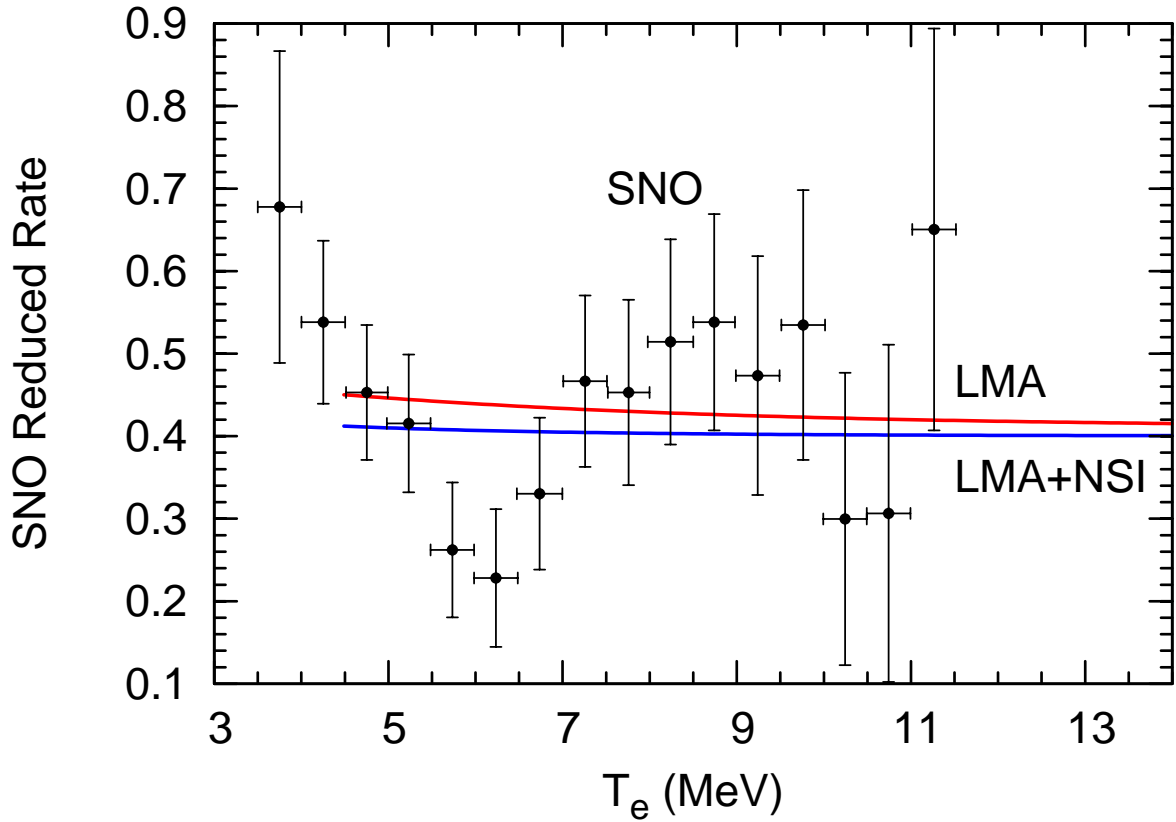}
\caption{\it Predictions for SNO neutrino electron scattering  
superimposed on the data \cite{Aharmim:2009gd} (units in MeV for electron kinetic 
energy). The upper curve is the LMA prediction and the lower curve is the LMA one 
with non standard interactions. Error bars are larger than in SuperKamiokande so 
that the data are consistent with a flat spectrum.}
\label{fig4}
\end{figure}

\noindent All cases in the first column of table \ref{table1} along with D2, E2, F2 in 
the second column lead qualitatively to the same monotonically decreasing probability curve: 
a high probability ($P\geq P_{LMA}$) for low energy ($E\lesssim 3~MeV$) and a low one 
($P\leq P_{LMA}$) for intermediate and high energies. The curve becomes  
increasingly flat in this energy sector as $\varepsilon_{\alpha\alpha}$ increases, 
which is also reflected in the flatness of the SuperKamiokande spectral rate.
However for cases A1, B1, C1 and D2, E2, F2 the probability gets too high for low 
energies so that the Ga \cite{Cattadori:2005gq,Gavrin:2007wc} rate fails to be 
conveniently fitted. The `best' results in the sense that they 
lead to the most flat spectral rate which approaches the SuperKamiokande one and 
to a correct fit for Ga are obtained alternatively from cases D1, E1, or F1 for the 
following values

\be
\begin{array}{ccccc}
\varepsilon_{\mu\mu}^{e~P}&=&-i|\varepsilon_{\mu\mu}^{e~P}| & = & -i~1.5\times 10^{-3}\\
\varepsilon_{\mu\mu}^{u~P}&=&-i|\varepsilon_{\mu\mu}^{u~P}| & = & -i~2.5\times 10^{-3}\\
\varepsilon_{\mu\mu}^{d~P}&=&+i|\varepsilon_{\mu\mu}^{d~P}| & = & +i~2.0\times 10^{-3}  .
\end{array}
\label{eq20}
\ee
For larger values of the NSI couplings the probability moves further away from its LMA profile 
so that the Ga rate becomes too high and the $^8 B$ one too low. In fig.\ref{fig1} we plot several
survival probabilities: the dashed line is the vacuum one, then
at the lowest energy and from bottom to top the first curve is the
LMA one, the next corresponds to all cases in (\ref{eq20}) and leads to the best fit of the four, the 
next one to the case 
$\varepsilon_{\mu\mu}^{e~P}=-i~3\times 10^{-3}$ and the top one to $\varepsilon_{ee}^{e~P}=
+i~5\times 10^{-3}$. A comparison is shown in table \ref{table2} between the predictions and the quality
of the fits obtained from LMA and the case $\varepsilon_{\mu\mu}^{e~P}=-i~1.5\times 10^{-3}$.
In figs.\ref{fig2} and \ref{fig3} we show the SuperKamiokande spectrum for LMA (upper curves) and for the first case 
in (\ref{eq20}) (lower curves) superimposed on the data points taken respectively from refs.
\cite{Fukuda:2002pe} and \cite{:2008zn}. The improvement obtained through the NSI coupling 
is clearly visible. In fig.\ref{fig4} the two curves are superimposed on 
the SNO data points for electron scattering \cite{Aharmim:2009gd}. Here the data are also 
clearly consistent with a constant rate.

\begin{table}[ht]
\centering
\begin{tabular}{|c|cccccccccc|} \hline \hline
     & Ga & Cl & SK & $\rm{SNO_{NC}}$ & $\rm{SNO_{CC}}$ & $\rm{SNO_{ES}}$ &
$\!\!\chi^2_{rates}\!\!$ & $\chi^2_{{SK}_{sp}}$ & $\chi^2_{SNO}$ & $\chi^2_{gl}$\\ \hline
LMA  & 64.9 & 2.84 & 2.40 & 5.47 & 1.79 & 2.37 & 0.67 & 42.0 & 48.6 & 91.3\\
$-i|\varepsilon_{\mu\mu}^{e~P}|$ & 69.7 & 2.74 & 2.23 & 5.47 & 
1.68 & 2.26 & 0.11 & 40.3 & 45.0 & 85.4\\ \hline
\end{tabular}
\caption{\it{Comparison between the LMA predictions for solar event rates and the NSI ones 
with $-i|\varepsilon_{\mu\mu}^{e~P}|=-i~1.5\times 10^{-3}$. For details of the $\chi^2$ analysis
see for instance \cite{Das:2009kw}.}}
\label{table2}
\end{table}

The first set of cases in the second column of table \ref{table1}, namely A2, B2, C2, lead qualitatively 
to the inverse behaviour with energy of the LMA probability. As $|\varepsilon_{\alpha\alpha}|$ 
increases from its lower bound, one gets $P\leq P_{LMA}$ for low energies ($E\lesssim 2-3~MeV$) 
and $P\geq P_{LMA}$ for intermediate and high energies, so that the fits worsen with respect to 
the LMA ones.

Furthermore all cases in the third column of table \ref{table1}, namely A3, B3, C3, D3, E3, F3 lead to
probability curves which are totally unsuitable: they deviate drastically from both $P_{LMA}$ 
and a from flat, suitable profile able to generate the SuperKamiokande spectrum.

We have also checked that combinations of real and imaginary parts for all couplings do 
not change the previous results. This should be expected since, as mentioned earlier, 
purely real couplings do not change the LMA probability. The only consequence of introducing
real parts in the NSI couplings comes in the spectral event rates through the quantities
$\tilde g_{L,R}$ in eqs.(\ref{eq14}) and (\ref{eq15}), but the differences lie much beyond the experimental
accuracy.

\subsection {$(\varepsilon_{\alpha \beta}^{eP})_{CC}\neq
(\varepsilon_{\alpha \beta}^{eP})_{NC}$}
\label{subsec3.2}

The analysis of the more general case of different CC and NC couplings affecting the 
$\nu_e~e$ scattering diagrams can be made quite simple if one examines the coefficients of 
$(\varepsilon_{\alpha \beta}^{eP})_{CC}$ and $(\varepsilon_{\alpha \beta}^{eP})_{NC}$ in
eq.(11). The first is unity whereas for equal CC and NC couplings it is 0.96 and the 
second is now -0.04 as compared to the previous value 0.96 as well. Consequently one 
expects that the analysis for $(\varepsilon_{\alpha \beta}^{eP})_{CC}$ leads to 
approximately the same results as for equal couplings while the results are modified by
a factor of 0.96/(-0.04) in the analysis for $(\varepsilon_{\alpha \beta}^{eP})_{NC}$.
Indeed the convenient modification in the LMA probability is obtained for 
\be
(\varepsilon_{\mu\mu}^{e~P})_{CC} =-i(|\varepsilon_{\mu\mu}^{e~P}|)_{CC} = -i~1.4\times 10^{-3}
\label{eq21}
\ee
or alternatively
\be
(\varepsilon^{e~P}_{\mu\mu})_{NC}=+i(|\varepsilon^{e~P}_{\mu\mu}|)_{NC}=+i~3.6\times 10^{-2}.
\label{eq22}
\ee
which lead to the same probability as the cases listed in eq.(\ref{eq20}). The results for the 
other couplings involving u and d quarks are of course unchanged. As before we have 
considered one coupling at a time to be non zero. 

\subsection{All diagonal couplings}
\label{subsec3.3}

We now allow for all diagonal entries of the NSI term in eq.(\ref{eq12}) to be nonzero and complex. 
In a first step to this generalization we start with all three couplings $\varepsilon_
{\mu\mu}^{e,u,d}$ nonzero simultaneously. Starting with the three equal in moduli, we find 
that the right modification of the LMA probability is obtained with

\be
\begin{array}{ccccc}
\varepsilon_{\mu\mu}^{e~P}&=&-i|\varepsilon_{\mu\mu}^{e~P}| & = & -i~0.7\times 10^{-3}\\
\varepsilon_{\mu\mu}^{u~P}&=&-i|\varepsilon_{\mu\mu}^{u~P}| & = & -i~0.7\times 10^{-3}\\
\varepsilon_{\mu\mu}^{d~P}&=&+i|\varepsilon_{\mu\mu}^{d~P}| & = & +i~0.7\times 10^{-3}~.
\end{array}
\label{eq23}
\ee

In a second step we allow for one diagonal entry in the NSI term of (\ref{eq12}) in addition to
$\varepsilon_{\mu\mu}^{e,u,d}$ to be nonzero and complex. We find that a finite $\varepsilon
_{\tau\tau}^{e,u,d}$ as added to the parameter choice (\ref{eq23}) leads to a `wrong' probability, 
unless $|\varepsilon_{\tau\tau}^{e,u,d}|\lesssim|(1/100)\varepsilon_{\mu\mu}^{e,u,d}|$
in which case the LMA solution remains unchanged. The same is true for $\varepsilon_{ee}^{e,u,d}
\neq 0$. However adding both $\varepsilon_{ee}^{e,u,d}$ and $\varepsilon_{\tau\tau}^{e,u,d}$ 
to (\ref{eq23}), the correct change in the LMA probability can be obtained provided 
$|\varepsilon_{ee}^{e,u,d}|$ and $|\varepsilon_{\tau\tau}^{e,u,d}|
\simeq |(1/10)\varepsilon_{\mu\mu}^{e,u,d}|$. To this end two choices are possible: 
either the signs of $\varepsilon_{ee}^{e,u,d}$ are changed with respect to those of table \ref{table1} 
with those of $\varepsilon_{\tau\tau}^{e,u,d}$ unchanged that is
\be
\varepsilon_{ee}^e=-i|\varepsilon_{ee}^e|,~\varepsilon_{ee}^u=-i|\varepsilon_{ee}^u|,~
\varepsilon_{ee}^d=+i|\varepsilon_{ee}^d|
\label{eq24}
\ee
and
\be
\varepsilon_{\tau\tau}^e=-i|\varepsilon_{\tau\tau}^e|,~\varepsilon_{\tau\tau}^u=
-i|\varepsilon_{\tau\tau}^u|,~\varepsilon_{\tau\tau}^d=+i|\varepsilon_{\tau\tau}^d|
\label{eq25}
\ee
or vice-versa.

A solution with all nine diagonal couplings having equal moduli is also possible. It corresponds
to 
\be
|\varepsilon_{ee}^{e,u,d}|\simeq|\varepsilon_{\mu\mu}^{e,u,d}|\simeq|\varepsilon_{\tau
\tau}^{e,u,d}|=(2-4)\times 10^{-4}~
\label{eq26}
\ee
provided the signs of $\varepsilon_{ee}^{e,u,d}$ are
unchanged with respect to table \ref{table1} and those of $\varepsilon_{\tau\tau}^{e,u,d}$ are reversed
that is
\be
\varepsilon_{ee}^e=+i|\varepsilon_{ee}^e|,~\varepsilon_{ee}^u=+i|\varepsilon_{ee}^u|,~
\varepsilon_{ee}^d=-i|\varepsilon_{ee}^d|
\label{eq27}
\ee
and
\be
\varepsilon_{\tau\tau}^e=+i|\varepsilon_{\tau\tau}^e|,~\varepsilon_{\tau\tau}^u=
+i|\varepsilon_{\tau\tau}^u|,~\varepsilon_{\tau\tau}^d=-i|\varepsilon_{\tau\tau}^d|~.
\label{eq28}
\ee
Notice that the reverse choice in signs with respect to (\ref{eq27}) and (\ref{eq28}) would leave the LMA
solution unchanged.

In this section we have analysed the conditions to be imposed on the NSI couplings in order
to obtain a suitable survival probability for a flat SuperKamiokande spectrum, an
improved Cl rate prediction relative to the LMA one and accurate predictions for all other
rates. As shall be seen next, these conditions while necessary, are not sufficient, owing
to the fact that neutrino decay in solar matter follows as a consequence of the imaginary 
diagonal couplings in the Hamiltonian.

\section{Neutrino decay in solar matter?}
\label{sec4}

We recall that a stable stationary state solution of the wave function, for a particle 
of energy $E$, contains a phase factor $e^{-iEt}$. If the Hamiltonian
has complex eigenvalues this phase factor becomes instead 
\be
e^{-i\left(E-i\Gamma\right)t}
\label{eq29}
\ee 
so the state is unstable, with decay rate $\Gamma$ and lifetime $\Gamma^{-1}$.

Since the suitable Hamiltonian that reduces the tension between the LMA solution and the 
data has imaginary diagonal couplings, its eigenvalues are complex which implies the 
existence of unstable states with a finite decay rate. We note that this instability 
is induced by the solar matter density [see eq.(\ref{eq12})] and in its absence, in the vacuum, the 
neutrinos are stable. Furthermore, as the number of neutrinos (or neutrinos and antineutrinos)
must remain constant as a consequence of unitarity, we must look for those solutions of the 
wave equation in which at least one but no more than two eigenvalues have a negative imaginary 
part. The remainder will have positive imaginary parts. This will ensure that the unstable state 
or states decay into any of the lighter. The negative imaginary parts correspond to the 
decaying states as it follows from (\ref{eq29}), while the positive ones to the states into 
which the former decay. Such criteria must be imposed together with the requirement of 
consistency with the solar neutrino data, namely a flat SK spectrum and accurate 
Ga, Cl and SNO rates. To this end the correct solution is found to be
\begin{equation}
{\cal H_{NSI}}=G_F\sqrt{2}N_e\left[x_1\left(\begin{array}{ccc}
\frac{i}{2}\varepsilon &  &\\ & -i\varepsilon &\\ & & \frac{i}{2}\varepsilon \end{array} \right)
+x_2\left(\begin{array}{ccc} \frac{i}{2}\varepsilon & &\\ & -i\varepsilon &\\
&  & \frac{i}{2}\varepsilon \end{array} \right)+
x_3\left(\begin{array}{ccc} -\frac{i}{2}\varepsilon &  &\\ & i\varepsilon &\\ &  &
-\frac{i}{2}\varepsilon \end{array} \right)\right]
\label{eq30}
\end{equation}
where $\varepsilon=3.5\times 10^{-4}$. Here 
\be
x_1=\frac{1}{2}+2sin^2\theta_W,~x_2=1-\frac{8}{3}sin^2\theta_W+
\frac{N_n}{2N_e},~x_3=-\frac{1}{2}+\frac{2}{3}sin^2\theta_W-\frac{N_n}{N_e}~
\label{eq31}
\ee
Notice that (\ref{eq30}) is consistent with (\ref{eq26}), (\ref{eq27}) and (\ref{eq28}).

\begin{figure} [htb]
\centering
\includegraphics[height=200mm,keepaspectratio=true,angle=0]{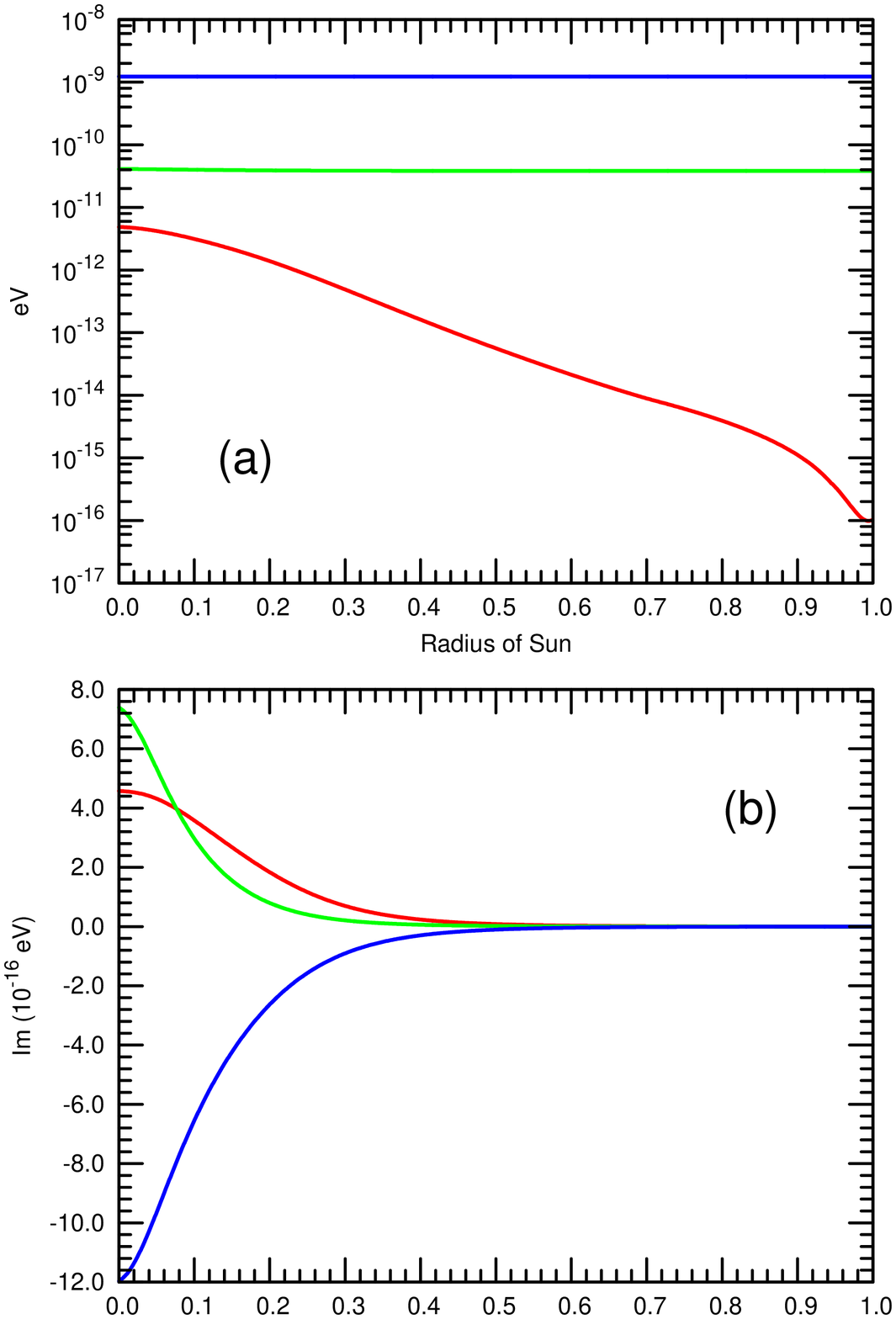}
\caption{\it The real (a) and imaginary parts (b) of the neutrino mass matter eigenvalues
for $E=1~MeV$: the lower curve in (b) associated to the decaying state corresponds to the 
upper curve in (a), the largest of the mass matter eigenvalues.}
\label{fig5}
\end{figure}

In fig.\ref{fig5} (panels (a) and (b)) we plot the real and imaginary parts of the eigenvalues 
of the Hamiltonian (\ref{eq12}) with the solution (\ref{eq30}) for the NSI term as a function of the 
solar fractional radius for a typical solar neutrino energy $E=1~MeV$. From panel (b) 
which displays the imaginary parts, it is seen that a significant decay of one of the 
mass matter eigenstates can occur during the first 30-40\% of the trajectory with an 
average decay rate $O(10^{-16}~eV)$. Further along the neutrinos appear to be stable 
while still in the sun and in the vacuum. As expected, the decaying state (the negative 
values in panel (b)) corresponds to the largest of the mass matter eigenstates (panel 
(a)).  

The neutrino decay modes considered in the literature involve photon or majoron emission
along with a neutrino or antineutrino. Although radiative decay is enhanced in a 
medium \cite{Giunti:1990pp}, for the sun such an enhancement is far too small, as it leads 
to a lifetime at least five orders of magnitude larger than the age of the universe, 
so such an effect is totally irrelevant here. The same holds for the `neutrino spin-light' 
in matter \cite{Lobanov:2004um,Grigoriev:2010uk}. We are therefore left with the
possibility of neutrino decay in matter into an antineutrino and a majoron $\chi$
\cite{Kachelriess:2000qc,Tomas:2001dh,Farzan:2002wx,Lessa:2007up}
\be
\nu_{i}\rightarrow\bar\nu_{j}~+~\chi
\label{eq32}
\ee
where subscripts $i$ and $j$ denote mass eigenstates. 

In the energy range of interest within the sun the neutrino transition is not matter dominated, hence 
mass and flavour eigenstates do not coincide. So, having started from the interaction potentials 
which are flavoured based, a rotation from flavour to mass states in solar matter is
needed in order to obtain the decay rate. However the quantity of interest which can be directly related 
with experiment is the antineutrino probability at the Earth for a particular flavour, namely $\bar\nu_e$. 
To this end the above mentioned rotation must be undone in the vacuum. Thus we will work from now on with 
flavour based quantities.

\begin{figure} [htb]
\centering
\includegraphics[height=195mm,keepaspectratio=true,angle=0]{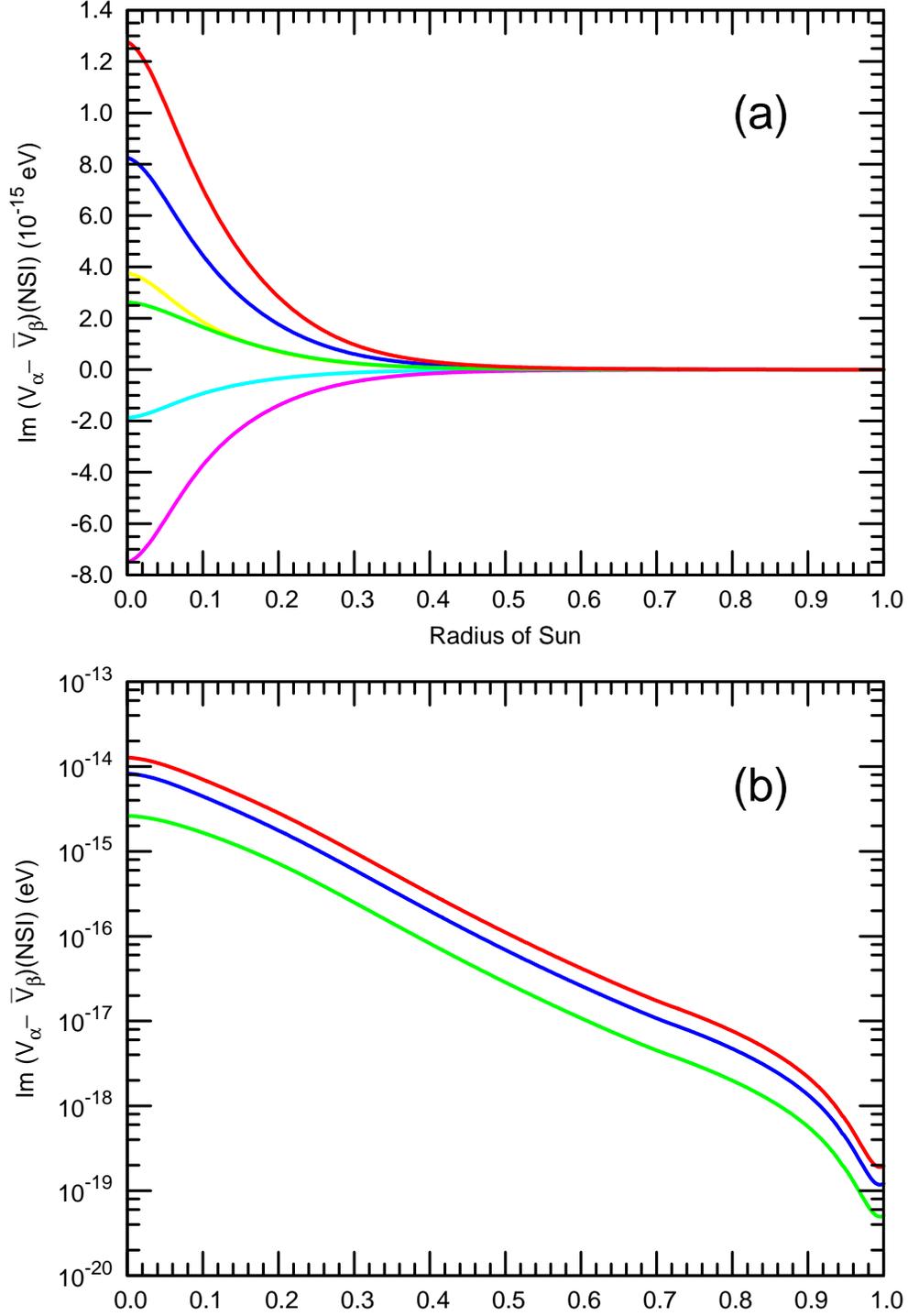}
\caption{\it Neutrino NSI potentials along the sun: panel (a) from top to bottom displays
the imaginary parts of $V_e-\overline V_e$, $V_e- \overline V_\tau$ (or $V_\tau-\overline V_e$), 
$V_{\tau}- \overline V_\tau$, $V_{e}-\overline V_\mu$ (or $V_\mu-\overline V_e$),
$V_{\mu}- \overline V_\tau$ (or $V_{\tau}-\overline V_\mu$)  and $V_{\mu}-\overline V_\mu$. 
The first, second and fourth from the top, pertaining to $\bar\nu_e$
production, are also shown in panel (b) in logarithmic scale.}
\label{fig6}
\end{figure}

The probability of antineutrino appearance for flavour $\beta$ per unit solar radius is then
\be
\frac{d P_{\bar\nu_{\beta}}(E_f)}{d r}=
\int_{E_f}^{E_{0_{max}}}\phi(E_0)~(1-e^{-\Gamma(r,E_0)r})~
\frac{d\Gamma(r,E_0,E_f)}{dE_f}~dE_0~.
\label{eq33}
\ee
Here $\phi(E_0)$ is the normalized $^8 B$ solar neutrino spectrum \cite{Bahcall} with $E_{0_{max}}=16.56~MeV$
and the factor $(1-e^{-\Gamma(r,E_f)r})$ takes into account the neutrino flux reduction. The quantity
$\frac{d\Gamma}{dE_f}$ denotes the differential decay rate per unit antineutrino energy $E_f$
\cite{Kachelriess:2000qc,Farzan:2002wx}
\be
\frac{d\Gamma}{dE_f}=\frac{|g_{\alpha\beta}|^2}{8\pi}\frac{E_0-E_f}{{E_0}^2}
|V_{\alpha}(r)-\overline V_\beta(r)|(NSI)
\label{eq34}
\ee
where $g_{\alpha\beta}$ are the neutrino-majoron couplings and 
$V_{\alpha}$, $\overline V_\beta$ the neutrino and antineutrino interaction potentials\footnote{Recall 
that neutrino and antineutrino interaction potentials are symmetric to each
other, $V_{\beta}=-\overline V_\beta$.}. 
The energy dependent antineutrino production probability from the decay (\ref{eq32}) is therefore 
\be
P_{\bar\nu_{\beta}}(E_f)\!=\!\frac{|g_{\alpha\beta}|^2}{8\pi}\!\!\int_{R_i}^{R_S}\!\!|V_{\alpha}(r)-\overline V_\beta(r)|(NSI)\!
\int_{E_{f}}^{E_{0_{max}}}\!\phi(E_0)(1-e^{-\Gamma(r,E_0)r})\frac{E_0-E_f}{{E_0}^2}~dE_0~dr
\label{eq35}
\ee
where $R_S$ is the solar radius and $R_i$ is the production point, assumed for simplicity
at 5\% of $R_S$ for $^8 B$ neutrinos. Upper bounds for the neutrino majoron couplings $g_{\alpha\beta}$ have
been estimated \cite{Kachelriess:2000qc,Lessa:2007up} and in particular the authors of ref. 
\cite{Lessa:2007up} find 
\be
\sum_{\alpha}|g_{e{\alpha}}|^2<5.5\times 10^{-6}
\label{eq36}
\ee
of particular interest here, since we will be concerned with $\bar\nu_e$ production for comparison
with the Borexino \cite{Bellini:2010gn} and KamLAND \cite{Eguchi:2003gg} upper bounds on solar $\bar\nu_e$.
The probability (\ref{eq35}) is calculated in the interval \cite{Farzan:2002wx} $[max\{1/2(V_{\alpha}-
\overline V_\beta),-\overline V_\beta\},E_{0_{max}}[$ and the decay (\ref{eq32}) requires 
$Im(V_{\alpha}-\overline V_\beta)>0$ with $V_{\alpha},~\overline V_\beta$ obtained from eqs.(\ref{eq11}), (\ref{eq30}) and (\ref{eq31})
\begin{eqnarray}
V_e(NSI) &=& \frac{i}{2}G_F\sqrt{2}(3.5\times 10^{-4})N_e\left(1-\frac{2}{3}sin^2\theta_W+\frac{3N_n}{4N_e}\right)\label{eq37}\\
V_{\mu}(NSI) &=& iG_F\sqrt{2}(3.5\times 10^{-4})N_e\left(-1+\frac{4}{3}sin^2\theta_W-\frac{3N_n}{2N_e}\right)\label{eq38}\\
V_{\tau}(NSI) &=& \frac{i}{2}G_F\sqrt{2}(3.5\times 10^{-4})N_e\left(1-\frac{4}{3}sin^2\theta_W+\frac{3N_n}{2N_e}\right)~.
\label{eq39}
\end{eqnarray}
We next evaluate the interaction potentials (\ref{eq37})-(\ref{eq39}) throughout the neutrino trajectory in the sun (see fig.\ref{fig6}), 
and analyse the possibilities for neutrino-antineutrino decay in terms of the necessary
condition $Im(V_{\alpha}-\overline V_\beta)>0$. To this end, inspection of fig.\ref{fig6} shows that the sums 
$V_e-\overline V_e$, $V_\tau-\overline V_e$ and ${V_\mu}-\overline V_e$
(the two top curves and the fourth from the top in panel (a)) contribute along the whole trajectory 
to $\bar\nu_e$ production. Similarly $\bar\nu_{\mu}$ production 
is less important, as it can depend only on $V_e-\overline V_\mu$ (fourth line from the top in (a)).
Finally $\bar\nu_{\tau}$ production 
comes from the contribution of $V_e- \overline V_\tau$ and $V_{\tau}- \overline V_\tau$ (second and third 
curves from the top in panel (a)) which are both positive over the whole trajectory.
The two bottom curves in panel (a) namely $V_{\mu}- \overline V_\tau$ (or $V_{\tau}-\overline V_\mu$) 
and $V_{\mu}-\overline V_\mu$ are both negative, they refer to 
$\nu_{\tau}$ and $\nu_{\mu}$ production and need not concern us here. To summarize, the production of:
\begin{itemize}
\item $\bar\nu_e$ involves $V_e-\overline V_e$, $V_\tau-\overline V_e$ and $V_{\mu}-\overline V_e$ 
\item $\bar\nu_{\mu}$ involves $V_e-\overline V_\mu$
\item $\bar\nu_{\tau}$ involves $V_e- \overline V_\tau$ and $V_{\tau}- \overline V_\tau$.
\end{itemize}

\begin{figure} [htb]
\centering
\includegraphics[height=200mm,keepaspectratio=true,angle=0]{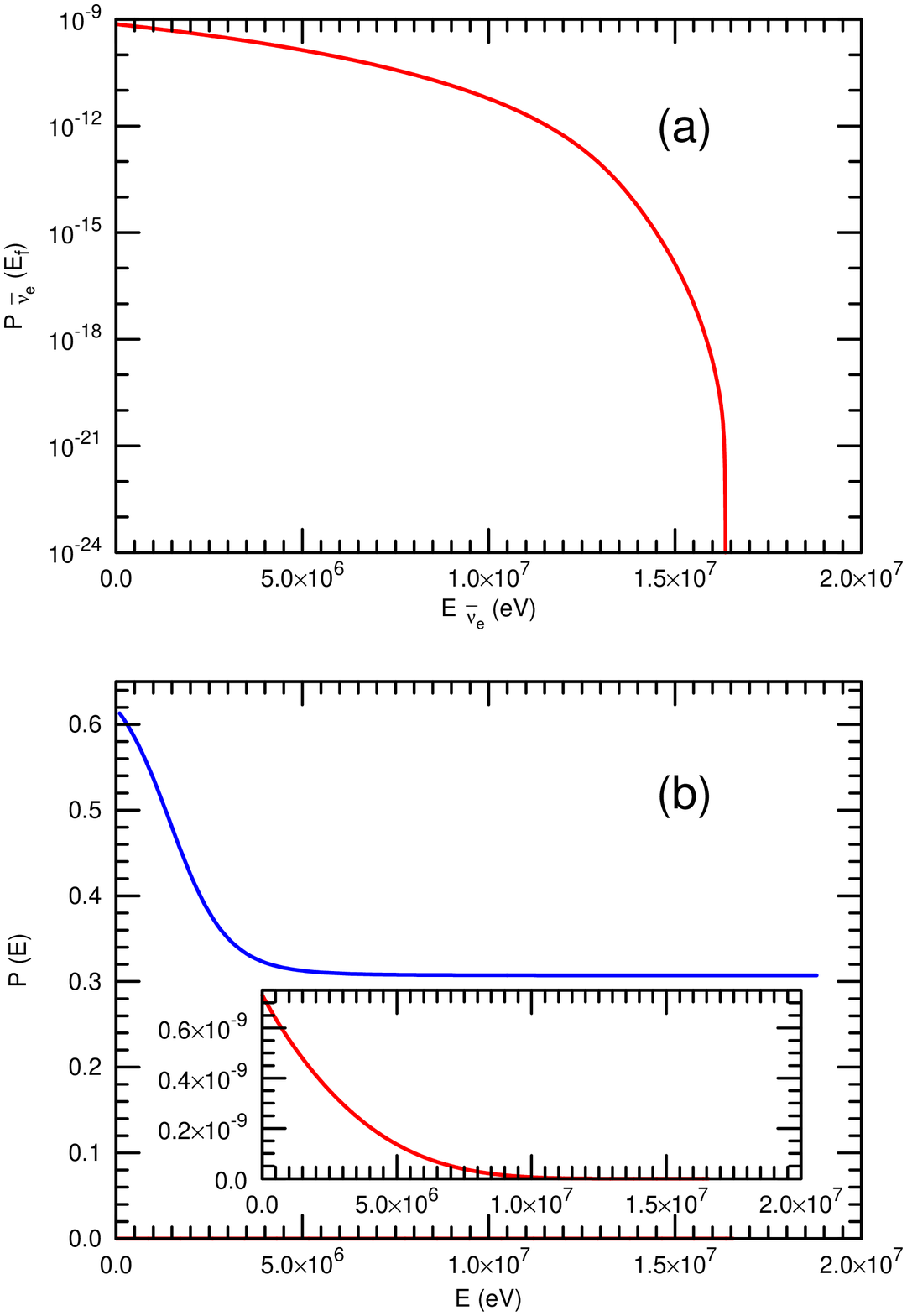}
\caption{\it Panel (a): antineutrino production probability in a logarithmic scale. Panel (b):  
LMA+NSI probability and antineutrino production probability (inner graph). Its extremely small
value prevents the possibility of observing antineutrinos from the sun from NSI.}
\label{fig7}
\end{figure}

Following these rules and using eq.(\ref{eq35}) with the upper limit (\ref{eq36}), we represent in fig.\ref{fig7} the upper bound of the
energy dependent $\bar\nu_e$ production probability, $P_{\bar\nu_e}(E_f)$. In panel (a) we use a logarithmic 
scale and in panel (b) both the LMA probability and $P_{\bar\nu_e}(E_f)$ (inner graph) 
are shown. The antineutrino probability is seen to be extremely small but grows rapidly as $E_f$ approaches 
its lower limit. It  
decrease fast to zero as $E_f\rightarrow E_{0_{max}}$, since fewer neutrinos contribute in the upper
energy range. 

In order to seek a comparison with the data, we first note that the Borexino upper bound on solar 
$\bar\nu_e$, namely $760~cm^{-2}s^{-1}$, applies for $E_{\bar\nu_e}>1.8~MeV$ which includes most
of the $^8 B$ spectrum \cite{Bellini:2010gn} and so corresponds in practice to a maximum total probability 
for $\bar\nu_e$ production \cite{PenaGaray:2008qe}  
\be
P_{\bar\nu_e}\leq\frac{760}{5.94\times 10^6}=1.3\times 10^{-4}.
\label{eq40}
\ee
We therefore consider the area limited by $P_{\bar\nu_e}(E_f)$ and the two extreme absciss{\ae} $1.8~MeV$ and 
$16.56~MeV$, dividing
it by the area limited by the unit probability and the same two absciss{\ae}. This ratio gives the
total probability for $\bar\nu_e$ production to be compared with (\ref{eq40}). We find for the ratio of these 
two areas  
\be
7.5\times 10^{-11}
\label{eq41}
\ee
which is within the Borexino bound by 6 orders of magnitude..

The KamLAND collaboration, on the other hand, reports a bound $370~cm^{-2}s^{-1}$ for $E_{\bar\nu_e}>8.3~MeV$
\cite{Eguchi:2003gg} which corresponds to a probability
\be
P_{\bar\nu_e}\leq\frac{370}{5.94\times 10^6~\int_{8.3}^{16.56}\phi(E)dE}=2.1\times 10^{-4}
\label{eq42}
\ee
whereas for the ratio of the corresponding areas we now get
\be
3.3\times 10^{-12}
\label{eq43}
\ee
within the KamLAND bound by 7-8 orders of magnitude. Hence any increase in experimental sensitivity will 
be unable to reveal a possible solar antineutrino flux produced from NSI.

\begin{figure} [htb]
\centering
\includegraphics[height=120mm,keepaspectratio=true,angle=0]{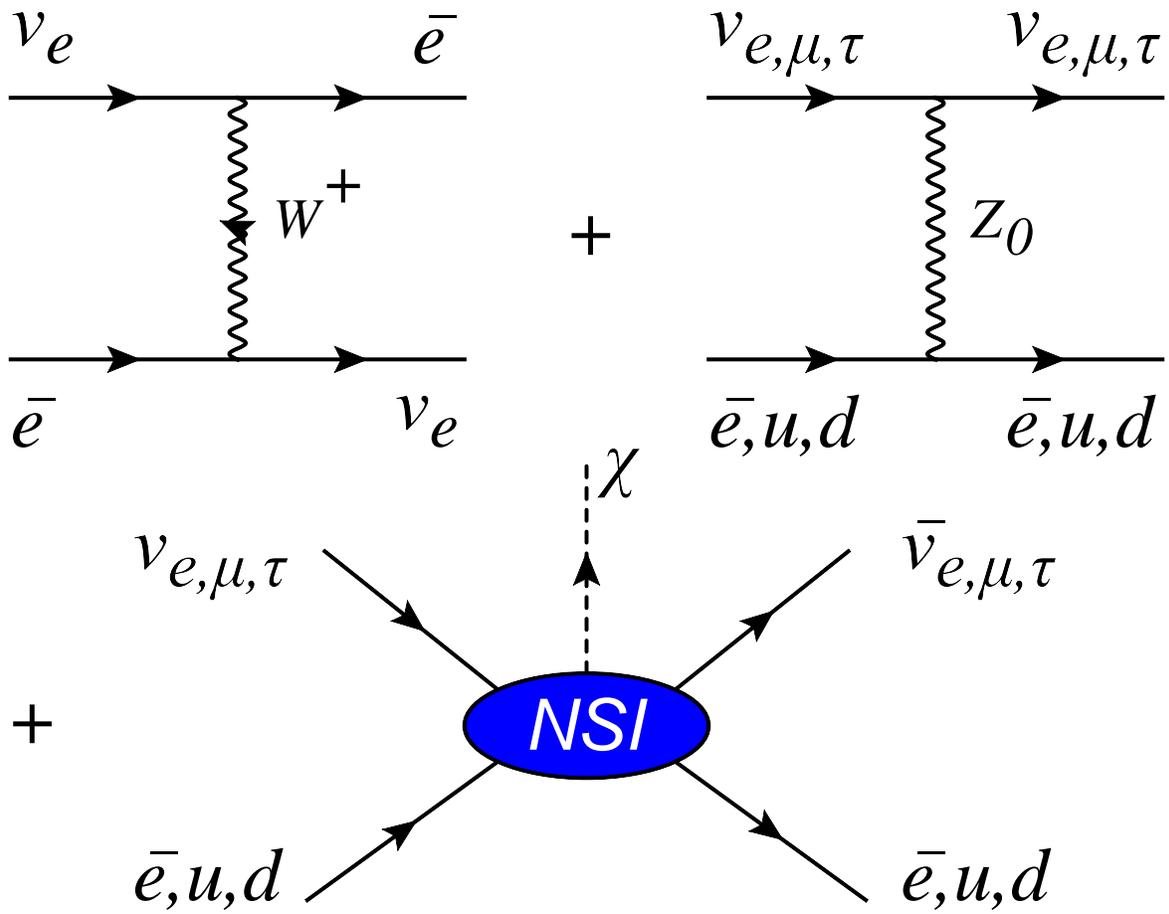}
\caption{\it The processes involved in the propagation of neutrinos in the sun as
in our model: the two upper diagrams are the standard ones for matter oscillation and 
the lower one represents the decay $\nu_i\rightarrow \bar\nu_{j}+majoron (\chi)$.}
\label{fig8}
\end{figure}

Finally the full physical process in our model for neutrino propagation and 
decay through NSI in the sun is represented in fig.\ref{fig8}. Equation (\ref{eq12}) we used for the Hamiltonian
does not take into account the extra physics involved in the majoron coupling and is 
therefore a truncated Hamiltonian, whose hermiticity is restored once the detailed majoron
emission process is taken into account.

\section{Conclusions}
\label{sec5}

We have investigated the prospects for improving the LMA predictions for solar neutrino
event rates with NSI. At present there is no evidence of any new physics associated to a 
scale not too far above the electroweak scale, hence the great variety of theoretical models 
available for NSI. In our approach we assumed that NSI are extra contributions 
to the vertices $\nu_{\alpha}\nu_{\beta}$ and $\nu_{\alpha}e$, so the new couplings describe 
the deviation from the standard model. With this in mind we derived the neutrino interaction 
potential in solar matter which was added to the standard Hamiltonian, proceeding with the
integration of the evolution equation through the Runge-Kutta method. Neutral and charged
current couplings are involved in interactions with electrons whereas only neutral couplings
affect those with quarks. We considered the new interactions both at the propagation and at 
the detection level. The improvement we searched for the LMA predictions consisted in 
finding whether and how the modification induced by NSI can lead to a flat spectral event 
rate for SuperKamiokande and an event rate for the Cl experiment within 1$\sigma$ of the 
data, while keeping the accuracy of all other predictions.

We used the current notation for the NSI couplings $\varepsilon_{\alpha\beta}^{e,u,d~P}$ where
$\alpha,\beta$ are the neutrino labels, $e,u,d$ denote the charged fermion involved in the
process and we investigated the range $|\varepsilon_{\alpha\beta}|~\epsilon~[5\times 
10^{-5}~,~5\times 10^{-2}]$. Our most remarkable and intriguing finding is that only the 
imaginary diagonal couplings in the Hamiltonian provide a solution to the tension between 
the LMA prediction for the SuperKamiokande spectrum and the data. This implies the
existence of unstable neutrino states, the instability being induced by the solar matter.
In the vacuum the neutrinos remain stable. 

We may now summarize our main results as follows

\begin{itemize}
\item Diagonal, imaginary couplings $\varepsilon_{\alpha\alpha}^{e,u,d~P}=
\pm i |\varepsilon_{\alpha\alpha}^{e,u,d~P}|$ are the only ones that lead to changes of all
kinds in the LMA probability and hence the rates.

\item Real couplings $\varepsilon_{\alpha\beta}^{e,u,d~P}=\pm|\varepsilon_{\alpha\beta}
^{e,u,d~P}|$ do not change the LMA probability when considered either one at a time or 
altogether and thus they induce a small change in the neutrino electron scattering rate 
($\lesssim 1\%$) which is far beyond experimental visibility. For the same reason the
addition of real parts to the imaginary couplings does not change the results.

\item Off diagonal couplings $\varepsilon_{\alpha\beta}^{e,u,d~P}$ ($\alpha\neq \beta$) whether
real or imaginary considered either one at a time or altogether do not change the LMA probability 
and thus the rates in a significant way.

\item Solving the tension between the LMA solution and the data, in particular predicting a
flat SuperKamiokande spectrum, requires unstable neutrinos, owing to the fact that only
imaginary diagonal couplings in the NSI Hamiltonian modify LMA. The requirement of unitarity 
then gives the following solution
\be
\begin{array}{cccccc}
\varepsilon_{ee}^{e,u}\!\!&\!\!=\!\!&\frac{i}{2}\varepsilon~~&~~
\varepsilon_{ee}^{d}\!\!&\!\!=\!\!&-\frac{i}{2}\varepsilon\\
\varepsilon_{\mu\mu}^{e,u}\!\!&\!\!=\!\!&-i\varepsilon~~&~~\varepsilon_{\mu\mu}^{d}\!\!&\!\!=& i\varepsilon\\
\varepsilon_{\tau\tau}^{e,u}\!\!&\!\!=\!\!&\frac{i}{2}\varepsilon~~&~~
\varepsilon_{\tau\tau}^{d}\!\!&\!\!=& -\frac{i}{2}\varepsilon
\end{array}
\label{eq44}
\ee
with $\varepsilon=3.5\times 10^{-4}~.$

\item The above facts imply neutrino decay from the heavier to the lighter states in the solar
matter probably with majoron emission, since radiative decay is irrelevant in the sun.
Having calculated the antineutrino appearance probability, we find however that it is quite 
unlikely to ever observe experimentally antineutrinos from the sun due to NSI. 
Small diagonal imaginary NSI couplings lead to quite a small and unobservable
antineutrino production probability. However this induces a visible and remarkable
change in both the survival and conversion probabilities to $\nu_{\mu}$ and $\nu_{\tau}$.

\item Experimentally the flatness of the SuperKamiokande and SNO electron energy spectra is 
at present the only evidence in favour of neutrino NSI in the sun.

\end{itemize}

Finally the detailed physics involved in majoron emission was not taken into account and the 
Hamiltonian we used eq.(\ref{eq12}) is therefore a truncated one which is sufficient for our purpose.

\section*{Acknowledgments}

{\em We are grateful to Marco Picariello, Lu\'{\i}s Lavoura and Sergio 
Palomares-Ruiz for useful discussions.
C. R. Das gratefully acknowledges a scholarship from Funda\c{c}\~{a}o para
a Ci\^{e}ncia e a Tecnologia (FCT, Portugal) ref. SFRH/BPD/41091/2007. This work was partially 
supported by the Marie Curie RTN MRTN-CT-2006-035505 and by Funda\c{c}\~{a}o para
a Ci\^{e}ncia e a Tecnologia through the projects
CERN/FP/ 109305/2009,  PTDC/FIS/098188/2008
and CFTP-FCT UNIT 777  which are partially funded through POCTI
(FEDER).}

\end{document}